\documentclass{llncs}
\usepackage{}
\usepackage{amsmath}
\usepackage{cases}
\usepackage{amsfonts}
\usepackage{amssymb}
\usepackage{makeidx}  % allows for indexgeneration
\usepackage{amsfonts,amssymb}
\usepackage{graphicx}
\begin{document}

\title{Approximation Resistance by Disguising Biased Distributions}
%\subtitle{[extended abstract]}
%
\titlerunning{Approximation Resistance}  % abbreviated title (for running head)
%                                     also used for the TOC unless
%                                     \toctitle is used
\renewcommand\thefootnote{}

\author{Peng Cui}

\authorrunning{Peng Cui}   % abbreviated author list (for running head)
%
%%%% modified list of authors for the TOC (add the affiliations)
\tocauthor{Peng Cui}
\institute{Key Laboratory of Data Engineering and Knowledge Engineering, MOE,
School of Information Resource Management, Renmin University of China, Beijing
100872, P. R. China.\\
\email{cuipeng@ruc.edu.cn}}

\maketitle              % typeset the title of the contribution

\begin{abstract}
In this short note, the author shows that the gap problem of some 3-XOR is NP-hard and can be solved by running Charikar\&Wirth's SDP algorithm for two rounds. To conclude, the author proves that $P=NP$.
\end{abstract}

\section{Introduction}
Max $k$-CSP is the task of satisfying the maximum fraction of constraints when each constraint involves $k$ variables, and each constraint accepts the same collection $C\subseteq G^k$ of local assignments. A challenging question is to identify constraint satisfaction problems (CSPs) that are extremely hard to approximate, so much so that they are NP-hard to approximate better than just outputting a random assignment. Such CSPs are called approximation resistant, including Max 3-SAT and Max 3-XOR as famous examples\cite{h}. A lot is known about such CSPs of arity at most four\cite{ha}, but for CSPs of higher arity, results have been scattered.

To make progress, conditional results are obtained assuming the Unique Games Conjecture (UGC) posed in \cite{k}. Under UGC, \cite{am} shows that a CSP is approximation resistant if the support of its predicate is the ground of a balanced pairwise independent distribution. However, the UGC remains uncertain, and it is desirable to look for new hardness reduction techniques. In \cite{ah}, the authors investigate $k$-CSP with no negations of variables and prove such $k$-CSP with the support of its predicate the ground of a biased pairwise independent distribution or uniformly positive correlated distribution or is approximation resistant in biased sense under Unique Games Conjecture.

In a recent work\cite{ch}, Chan obtains a general criterion for approximation resistance of the NP-hardness of Max $k$-CSP. He shows hardness for CSPs satisfying the support of its predicate $C\subseteq G^k$ is a subgroup and the uniform distribution over $C$ is a balanced pairwise independent distribution, where the domain is an Abelian group $G$. A random assignment satisfies $|C|/|G|^k$ fraction of constraints in expectation. His hardness ratio is tight up to an arbitrarily small constant under the standard assumption $P\ne NP$.

In his work, Chan views a Max $k$-CSP instance as a $k$-player game, and reduces soundness by a technique called direct sum. Direct sum is like parallel repetition, aiming to reduce soundness by asking each player multiple questions at once. However, with direct sum each player gives only a single answer, namely the product of answers to individual questions. His work borrows the idea of blocking distribution from \cite{ow}, which proves a new point of NP-hardness of Unique Games Problem using Moshkovitz and Raz Theorem\cite{mr}, other than the point of NP-hardness implied by the work of \cite{h}.

Unable to decrease soundness directly, he instead demonstrates randomness of replies. The crucial observation is that correlation never increases with direct sum. It remains to show that, in the soundness case of a single game, he can isolate any player of his choice, so that the player's reply becomes uncorrelated with the other $k-1$ replies after secret shifting. Then the direct sum of $k$ different games will isolate all players one by one, eliminating any correlation in their shifted replies. He proves the main result using the canonical composition technique. In the soundness analysis of the dictatorship test, he invoke an invariance-style theorem, based on \cite{ow}. Note that direct sum is in fact not necessary in the case of $k=3$.

The following is from the main theorem in \cite{ch}.

\begin{theorem}
For arbitrarily small constant $\varepsilon$, it is NP-hard to distinguish the following two cases given an instance $P$ of 3-XOR:

\begin{itemize}
\item Completeness: $\mathrm{val}(P)\ge 1-\varepsilon$.
\item Soundness: $\mathrm{val}(P)\le\frac{1}{2}+\varepsilon$.
\end{itemize}

\end{theorem}

In this short note, the author shows that the gap problem of folded 3-XOR as in the statement of Theorem 1 is can be solved by running Charikar\&Wirth's SDP algorithm\cite{cw} for two rounds. This leads to the fact that 3-SAT can be solved by an algorithm in polynomial time. Thus, the author settles the longstanding open problem in computational complexity theory, i.e., P vs NP problem.

\begin{theorem}
$P=NP$.
\end{theorem}

This work has an origin that conditionally strengthens the previous known hardness for approximating Min 2-Lin-2 and Min Bisection, assuming a claim that refuting Unbalanced Max 3-XOR under biased assignments is hard on average\cite{c}. In this paper, the author defines "bias" to be a parameter of pairwise independent distribution, while he defines "bias" to be the fraction of variables assigned to value 1 in \cite{c}. The author notices that biased pairwise independent distribution is defined in \cite{am,ah} and uniformly positively correlated distribution is defined in \cite{ah}.

\section{Definitions}
As usual, let $[q]=\{1,2,\cdots,q\}$, and $-[q]=\{-q,-q+1,\cdots,-1\}$.

Let $G=\{1,-1\}$, here 1 represent "0/false" and -1 represent "1/true" in standard Boolean algebra.

Denote the set of probability distributions over $G$ by

$$\bigtriangleup_{G}\triangleq\{x\in\mathbb{R}_{\ge 0}^{|G|}|\parallel x\parallel_{\ell^1}=1\}.$$

Random variables are denoted by italic boldface letters, such as $\vec z$. Suppose $\varphi$ is a distribution over $G^k$, the ground of $\varphi$ is defined as

$$G_\varphi=\{\varphi(\vec z)>0|\vec z\in G^k\}.$$

\begin{definition}
For some $0<\gamma<1$, a distribution $\varphi$ over $G^k$ is biased pairwise independent if for every coordinate $i\in[k]$,

$$\mathbb{P}[\vec z_i=1]=\gamma$$

\noindent and for every two distinct coordinates $i,j\in[k]$,

$$\mathbb{P}[\vec z_i=1,\vec z_j=1]=\gamma^2,$$

\noindent where $\vec z$ is a random element drawn by $\varphi$. $\gamma$ is called bias of $\varphi$. If $\gamma=\frac{1}{2}$, we say $\varphi$ is balanced pairwise independent.
\end{definition}

The author notices that a distribution over $G^k$ can be thought as a linear superposition of several distributions over $G^k$.

\begin{definition}
Given $m$ distributions $\varphi_l$ over $G^k$ with disjoint grounds $G_{\varphi_l}$, let $\psi$ is a distribution over $[m]$ with $\psi_l>0$ for each $l\in[m]$, and $\varphi$ be the distribution over $G^k$ such that

$$\varphi(\vec z)=\sum^{m}_{l=1}\psi_l\varphi_l(\vec z),$$

\noindent for each $\vec z\in G^k$. We say $\varphi_l$'s are disguised by $\psi$ to $\varphi$.
\end{definition}

\section{Dictatorship Test}
Theorem 1 is based on a dictatorship test $T$ satisfying the desired completeness and soundness properties.

The instance of Label-Cover $L$ (cf. \cite{mr})is a bi-regular graph $((U,V),E)$ with two parameters $d=2^{\mathrm{poly}(\frac{1}{\sigma})}$ and $R=\mathrm{poly}(\frac{1}{\sigma})$, where $\sigma$ is an arbitrarily small positive. Vertices from $U$ are variables with domain $[R]$ and vertices from $V$ are variables with domain $[dR]$. Every edge $e=(u,v)$ is associated with a map $\pi_{e}$, also denoted as $\pi_{u,v}$, where $\pi_{e}:[dR]\rightarrow[R]$ satisfying $|\pi^{-1}_e(t)|=d$ for each $t\in[R]$. Given an assignment $A:U\rightarrow[R],V\rightarrow[dR]$, $e$ is satisfied if $\pi_{e}(A(v))=\pi_{e}(A(u))$. The goal of $L$ is to seek an assignment to maximize the number of satisfied edges. An assignment that satisfies every edge is called {\it perfect assignment}.

As in \cite{ch}, we compose a $3$-player dictatorship test with a Label-Cover instance, which is a game involving the variable party and the clause party. Before composition, one player in the variable party replies over alphabet $[R]$ and all other players in the clause party reply over alphabet $[dR]$. Both alphabets are partitioned into $R$ blocks, each of which has size 1 for the variable party and size $d$ for the clause party. The $t$-th block is $\pi^{-1}_e(t)$. After composition, the players reply over domain $G$. We single out player $1$ as the lonely player, who is in the variable party, players $2$ and $3$ are in the clause party.

For every edge $e=(u,v)$ in $E$, a $3$-player $1$-lonely $C$-test $T$ is a $3$-tuple of random variables

$$\vec z=(\vec z^{(1)},\vec z^{(2)},\vec z^{(3)})\in G^{R}\times G^{dR}\times G^{dR}.$$

\noindent $C$ is the ground of a balanced pairwise independent distribution $\varphi$. The $C$-constraint associated with $e$ consists an assignment $f_e=(f_{1,v_1},f_{2,v_2},f_{3,v_3})$ to variables $v_i$'s, where $v_{1,1}=u$, and $v_{1,[2,R+1]}=\vec z^{(1)}$, $v_{i,1}=v$, and $v_{i,[2,dR+1]}=\vec z^{(i)}$ for $i=2,3$.

We think of $\vec z$ as an $R\times 3$ matrix, where columns are $\vec z^{(i)}$'s for $i\in[3]$ and entries are from $G$ in first column and are from $G^d$ in other columns. $\vec z$ is drawn from distribution $\mu$ determined as follows: For each row $t\in[R]$, independently choose $3$-tuples from $C$ by $\varphi$ for $d$ times as $\vec z^{(i)}_t$, agreeing at column $1$. By the construction of the dictatorship test, $z^{(1)}$ is uniformly random over $G^{R}$, and $z^{(i)}$ for $i=2,3$ is uniformly random over $G^{dR}$. Since $\varphi$ is balanced pairwise independent, looking at a column $1$ and any other column $i$ for each row, the marginal distribution is pairwise independent over $G\times G^d$, and looking at two columns $2$ and $3$ for each row, the marginal distribution is pairwise independent over $G^d\times G^d$.

Inspired by \cite{ow} and \cite{ch}, we also consider an uncorrelated version of the distribution $\mu$, $\mu'$, and an uncorrelated version of the test $T$, $T'$, in our analysis. Let $\mu$ be the distribution defined above. The partially uncorrelated distribution $\mu'$ is defined as: A matrix from $\mu'$ is chosen exactly as in $\mu$, and then column $1$ is re-randomized to be a uniformly random element from $G^R$.

Given an instance $L$ of Label-Cover, our reduction from $L$ to Max $C$ produces an instance that is a $k$-partite hypergraph on the vertex set $V_1\cup\cdots\cup V_k$. The first vertex sets $V_1$ is $V\times G^{R}$, and the other two vertex sets $V_i$ is $V\times G^{dR}$. All vertices are variables with domain $G$. For a $v_i\in V_i$, we write the first component of $v_i$ as $v_{i,1}$, the remaining components of $v_i$ as $v_{i,[2,dR+1]}$.

We think an assignment $f_{1,v_1}$ to variables in $v_1\in V_1$ as a function

$$f_{1,v_{1,1}}:G^{R}\rightarrow\Delta_G, v_{1,[2,R+1]}\mapsto f_{1,v_{1,1}}(v_{1,[2,R+1]}).$$

\noindent and an assignment $f_{i,v_i}$ to variables in $v_i\in V_i$ for $i=2,3$ as a function

$$f_{i,v_{i,1}}:G^{dR}\rightarrow\Delta_G, v_{i,[2,dR+1]}\mapsto f_{i,v_{i,1}}(v_{i,[2,dR+1]}).$$

For every edge $e=(u,v)$, the reduction introduces weighted $C$-constraints on the folded and $\eta$-noisy assignments $f_{i,v_i}$, as specified by the dictatorship test $T$.

\section{Proof of Theorem 2}
Let $G_m$ denote the subset of $G^3$ including all $3$-tuples with exactly $m$ $1$.

Suppose $C=G_3\cup G_1$, $\psi=(\frac{3}{4},\frac{1}{4})$, then $C$ is a subgroup of $G^3$ and the uniform distributions over $G_3$ and over $G_1$ are are disguised by $\psi$ to a balanced pairwise independent distribution. The Fourier spectra of $C$ is $C(y)=\frac{1}{2}+\frac{1}{2}{y_1y_2y_3}$. Let $P^{(3)}(y)$ be the tri-linear term in the Fourier spectra of $C$, then $P^{(3)}(y)=\frac{1}{2}$ for any $y\in C$.

In the dictatorship test,

\begin{equation*}
\begin{split}
&\mathbb{P}[\vec z_1=1,\vec z_2=-1,\vec z_3=-1]=\mathbb{P}[\vec z_1=-1,\vec z_2=1,\vec z_3=-1] \\
&=\mathbb{P}[\vec z_1=-1,\vec z_2=-1,\vec z_3=1]=\mathbb{P}[\vec z_1=1,\vec z_2=1,\vec z_3=1] \\
&=\textstyle\frac{1}{4},
\end{split}
\end{equation*}

\noindent and

\begin{equation*}
\begin{split}
&\mathbb{P}[\vec z_1=-1,\vec z_2=1,\vec z_3=1]=\mathbb{P}[\vec z_1=1,\vec z_2=-1,\vec z_3=1] \\
&=\mathbb{P}[\vec z_1=1,\vec z_2=1,\vec z_3=-1]=\mathbb{P}[\vec z_1=-1,\vec z_2=-1,\vec z_3=-1] \\
&=0.
\end{split}
\end{equation*}

\noindent where $\vec z$ is a random element drawn by $\varphi$.

Given an instance $P$ of Max $C$, by Theorem 1, for arbitrarily small constant $\varepsilon$, it is NP-hard to distinguish the following two cases: $\mathrm{val}(P)\ge 1-\varepsilon$; $\mathrm{val}(P)\le\frac{1}{2}+\varepsilon$.

On the other hand, let $I^{(3)}$ be the sum of tri-linear terms in the Fourier spectra of $P$. Suppose $\mathrm{val}(P)\ge 1-\varepsilon$ for some $\varepsilon$, there is an assignment $f^*$ under which is at least $I^{(3)}\ge\Omega(1)$ (cf. Lemma 4 in \cite{ha}).

Let $I^{(2)}$ be the sum of bi-linear terms defined as: For each tri-linear term $a_{i_1i_2i_3}x^{(1)}_{i_1}x^{(2)}_{i_2}x^{(3)}_{i_3}$ in $I^{(3)}$, introduce a bi-linear term $a_{i_1i_2i_3}x^{(1)}_{i_1}x^{(23)}_{i_2i_3}$, where $x^{(23)}_{i_2i_3}$'s are new variables in $G$, where $i_1\in[M]$ and $i_2,i_3\in[N]$.

Run Charikar\&Wirth's SDP algorithm\cite{cw} for two rounds as follows:

\begin{itemize}
\item Step 1, run Charikar\&Wirth's SDP algorithm for the first round on $I^{(2)}$ to get an assignment $f^{(1)}$ on $x^{(1)}_{i_1}$'s and $x^{(23)}_{i_2i_3}$'s.
\item Step 2, run Charikar\&Wirth's SDP algorithm for the second round on $I^{(3)}$ subject to $f^{(1)}$ to get an assignment $f^{(2)}$ to $x^{(2)}_{i_2}$'s and $x^{(3)}_{i_3}$'s.
\item Step 3, let $f:=f^{(1)}$ for $x^{(1)}_{i_1}$'s and let $f:=f^{(2)}$ for $x^{(2)}_{i_2}$'s and $x^{(3)}_{i_3}$'s.
\end{itemize}

The first round returns $f^{(1)}$ under which $I^{(2)}$ is at least $\Omega(1)$ (cf. Lemma 5 in \cite{cw}). By enumeration arguments, there is an assignment $f'$ to $x^{(2)}_{i_2}$'s and $x^{(3)}_{i_3}$'s under which $I^{(3)}$ subject to $f^{(1)}$ is at least $\Omega(1)$. Hence the second round returns $f^{(2)}$ under which $I^{(3)}$ subject to $f^{(1)}$ is at least $\Omega(1)$ (cf. Lemma 5 in \cite{cw}).

Therefore, the algorithm returns a solution of $P$, $f$, with expected value at least $\frac{1}{2}+\Omega(1)$ (cf. Theorem 3 in \cite{ha}). The proof of Theorem 2 is accomplished.

\end{document}